\documentclass{elsart1p}

\usepackage{amssymb}

\begin{document}

\begin{frontmatter}



\title{ ZQCD and universal matching to domain walls}


\author{C.P. Korthals Altes}

\address{Centre Physique Th\'eorique, B.P.907\\
 Luminy, 13288 Marseille Cedex. }

\begin{abstract}
We discuss the one loop matching procedure in ZQCD. Universality and Casimir scaling leave-in terms of the 't Hooft coupling- just two combinations of parameters to be fixed numerically. These numbers are then the same for any number of colours.

\end{abstract}

\end{frontmatter}

\section{Introduction}
\label{intro}
A 3d effective theory of QCD incorporating the center group 
symmetry has long defied the practitioners of thermal QCD.

The challenge is to marry in the model the  non-perturbative nature of the centergroup~\cite{pisarski}
which involves large gauge transformations- and the perturbative nature
of EQCD at high temperature.   
Recently Vuorinen and Yaffe~\cite{vuoryaffe} have furnished such a theory. This theory has the colour-magnetic gauge degrees of freedom (with field strength $F_{ij}$), and as matter field an $N\times N$ complex scalar field $Z={L\over N}{\bf 1}+M$. The field M is traceless,
so $L=Tr Z$. The field Z is the Polyakov loop, but not anymore unitary.

Its unitarity is lost by averaging over space and integrating out the colour-electric modes of QCD. The field $M=H+iA$, with both H and A Hermitean and traceless, is a remnant of the
  latter  modes.
To garantuee the resurgence of the massless fourth component of the vector potential (in our case A) the theory is supposed to have a large global $SU(N)\times SU(N)$ invariance, broken to the diagonal SU(N), which becomes the 
gauged colour group. Thermal effects are given by a potential $V_1$ and screen the fourth component of the vector potential
so should break explicitely the global invariance but should obviously respect colour and Z(N) invariance. 
Transformation laws are

\vspace{0.1cm}

\begin{itemize}
\item Z(N) symmetry: $L\rightarrow zL, M\rightarrow zM$
\item $SU(N)\times SU(N)$  symmetry:~$ M\rightarrow UMV^{\dagger}$.
\end{itemize}

\vspace{0.1cm}

Such a model can be used on the lattice to extract information on the 
hot phase of QCD, down to $T_c$. To have a straightforward continuum limit one would like to have superrenormalizable operators.

Lattice calculations for any number of colours have taught us important features of QCD, so we want to keep N general.

This being said we can now write the action of the model:
\begin{eqnarray}
S_{ZQCD}&=&{1\over{g_3^2}}(Tr\vec (DZ)^{\dagger}.\vec DZ +V_0+V_1+Tr F_{ij}^2)\nonumber\\
V_0&=&-M_Z^2TrZ^{\dagger}Z+{\Lambda\over N} (TrZ^{\dagger}Z)^2+\tilde\Lambda Tr(Z^{\dagger}Z)^2 \nonumber\\
&+&\Lambda_d \det Z+c.c.
\label{vo}
\end{eqnarray} 
\noindent with $\vec D=\vec\partial +i[\vec A,$ and $F_{ij}=\partial_i A_j-\partial_j A_i+i[A_i,A_j]$. The dimensionful gauge coupling is $g_3$, and in this notation all fields and couplings have engineering dimensions  as in four dimensions.

If $M_Z^2$ is positive we can set $Z=Z_c{\bf 1}+g_3Z_q$, determine 
$Z_c$ by minimizing, and see that the field A is the Goldstone boson.
Ignoring $V_1$ one has  indeed zero mass for the field A and 
non-zero mass for all other fields. 

$V_1$ contains radiative effects  like Debye screening, so is ${\cal O}({g_3^2\over {M_Z}})$. It  can be split in a natural way 
into two parts $V_1=V_U+V_Z$. $V_U$ is superrenormalizable for all values of N. By inspection one sees $V_U$ is U(1) invariant, instead
of only Z(N) invariant.  The other part is $V_Z$ containing all other
possible operators which are colour singlet and Z(N) invariant.
Explicitly:
\begin{eqnarray}
V_U&=&m_M^2TrM^{\dagger}M+\lambda_M(TrM^{\dagger}M)^2
+\tilde\lambda_M Tr(M^{\dagger}M)^2\nonumber\\
&+&\lambda^{'}_M|TrM^2|^2+\tilde\lambda_M^{'}TrM^2{M^{\dagger}}^2\nonumber\\
&+&\lambda_{LM} Re L^2Tr(M^{\dagger})^2
+\lambda^{'}_{LM}Re L^*TrM^2M^{\dagger}
\label{vu}
\end{eqnarray}
\noindent
whilst
$V_Z=\nu_NReTrM^N+....$ contains no U(1) invariants, only Z(N) invariants, and
grows like the number of partitions of N, and is {\it non}-superrenormalizable for $N\ge 5$.

    The problem with this approach is a practical one already  in the case of N=3: it has many 
    unknown parameters in the potentials. The purpose of this note is to get more insight in the matching. In particular we note in 
    eq. (\ref{wallmatch})  that the effective potential giving rise to the domainwall 
shows, apart from Casimir scaling, independence of N.

\section{Matching with EQCD}\label{sec:eqcdmatch}
EQCD is obtained from QCD by integrating out all modes with
mass $\sim T$:
 
 \begin{eqnarray}
S_{EQCD}&=&{1\over {g_3^2}}(Tr (\vec DA_0)^2+m_E^2Tr A_0^2+\lambda_E(Tr(A_0^2)^2\nonumber\\
&+&\tilde \lambda_ETrA_0^4+Tr F_{ij}^2....),
\end{eqnarray}

The EQCD parameters are to O($g^4$): 
\begin{equation}
m_E^2={g^2NT^2\over 3},\lambda_E={g^2N\over{4\pi^2N}},\tilde\lambda_E={g^2N\over{12\pi^2}}.							\end{equation}											
To this order $m_E=m_D$, the Debye mass.								
  Matching to EQCD is done in the bulk vacuum by integrating out the heavy fields. Because of chiral invariance this gives only derivative terms. The matching is then immediate from $V_U$ (identifying A with $A_0$):

\begin{eqnarray}
g_3^2=g^2T\nonumber\\
m_M^2-\lambda_{LM}\tilde v^2=m_E^2\nonumber\\
 \lambda_M+\lambda'_M=\lambda_E\nonumber\\
\tilde\lambda_M+\tilde\lambda'_M=\tilde\lambda_E
\end{eqnarray}

\section{Matching to the walls}\label{sec:matching}
The wall between Z(N) bulk vacua is characterized by a profile that
interpolates the value of the Polyakov loop in one side and of the other 
side of the wall. What matters for the wall is only the phase difference
so we can set without loss of generality the phase equal 1 and $\exp(ik{2\pi\over N})$ in the two vacua.

First ignore $V_1$.  Due to $SU(N)\times SU(N)$ invariance  the interpolating field $Z_{w}$  can be written as ${v\over N}U$ with U an $SU(N)$ matrix, provided $-\Lambda_{d},\Lambda, \tilde\Lambda \ge 0$. $U={\bf 1}$ resp $\exp(ik{2\pi\over N})$ in the two vacua.
In perturbation theory  $v$ does not depend on where we are in the wall and equals the bulk VEV all the way through the wall. The gulley
leading from one bulk vacuum to the other is flat. This degeneracy is lifted through fluctuations we now discuss.

The wall profile depends on v and the eigenphases of U. That means it is a gauge independent quantity. To compute the wall and its free energy density we compute the path integral with the constraint that the eigenphases and v are fixed~\cite{korthals}. This gives us the gauge independent  
effective action $G(\{Z_{w}\})$, and finally we minimize it with the apropriate boundary conditions on both sides of the wall thus fixing  the wall profile $Z_w$.

There is a considerable simplification: we can choose

\begin{equation}
Z_{w}={v\over N}\exp(iqY_k{2\pi\over N})\nonumber.
\end{equation}
Here q is a variable that varies inside the wall from 0 to $2\pi$.
\begin{equation}
Y_k = {1 \over N} ~{\rm diag}(\underbrace{k,k,\dots,k}_{N-k~{\rm times}},
       \underbrace{k-N,k-N,\dots,k-N}_{k~{\rm times}})\nonumber. 
\end{equation}
is the generalized hypercharge and generates Z(N) elements (1 and 
$\exp ik{2\pi\over N}$)
 on both sides of the wall, characterizing the bulk vacuum. With this choice the wall profile $Z_w={L_w\over N}{\bf 1}+M_w$ with  $M_w\sim Y_k$.
 
 To compute the fluctuations around $Z_w$
one sets $Z=Z_w+g_3Z_q$ and $\vec A=g_3 \vec Q$. The terms linear in $Z_q$ are changed by the constraints  on the eigenphases in the path integral. This change applies also to the masses 
of the fluctuating fields, because there is always a quadratic constraint
on the eigenphases, like e.g. on $Tr Z^{\dagger}Z$.   Note that the mass of the transverse gauge field
is induced by the wall in the kinetic term of the Z field, and that it is of the same order as the masses of all the other fluctuation fields. 

  To get rid of bilinear couplings of $\vec Q$ and $M_q$ induced by the wall, put  $R_{\xi}$ gauge~\cite{deforkurk}:
\begin{equation}
{\cal{G}}_{gf}={1\over{\xi}}\big(\vec\partial.\vec Q+i{\xi \over 2}([M_w^{\dagger},M_q]+[M_w,M_q^{\dagger}])\big)^2.
\end{equation}

For N=2 (where $Z={\Sigma\over 2}+i\Pi$, $\Sigma$ real, $\Pi$ Hermitean and traceless) this gauge condition produces the mass of the $\Pi^{\pm}$ components 
 (due the gauge constraints) 
and  of course that of the ghost and longitudinal fluctuations. Their common mass is  $4\xi v^2\sin(q/2)^2$. So the corresponding  three fluctuation determinants  cancel and hence the $\xi$ dependence, as it should.

Only the determinant of the    transverse potential, and  eventually of  remaining components $\Sigma_q$ and $\Pi^3_q$ of the Z field do contribute. They do mix inside the wall. The physical states have only background independent 
masses, hence their bulk values 0 and $\Lambda v^2$ (neglecting ${\cal O}(g_3^2/M_Z)$ contributions from $V_1$). Apart from the determinant also $V_1(L_w, M_w)$ contributes to the effective potential.
One finds for the effective potential  in one loop QCD respectively one loop ZQCD ($z^{'}=m_Dz)$, using the results from section ~\ref{sec:eqcdmatch}:
 \begin{eqnarray}
 G(q)/k(N-k)T^3&=&{4\pi^2T\over{3m_D}}\int dz^{'}((\partial_{z^{'}}q)^2
 +q^2(1-q)^2)\nonumber\\
 G(q)/k(N-k)T^3&=&{4\pi^2T\over {3m_D}}\int dz^{'}((\partial_{z^{'}}s)^2
+s^2(1-s)^2
+c_ks^3+b_ks^4+z_ks^6+...)\nonumber\\
\label{wallmatch}
\end{eqnarray}
where $s\equiv {v\over NT}{\sin(\pi q)\over {\pi}}$ in the effective action for ZQCD. This is the main result.

\begin{itemize}
\item Note Casimir scaling and universality: no k and N dependence respectively on the r.h.s.. Hence in ZQCD matching means $b_k=b, c_k=z_k=0$.
\item Fluctuation determinants generate cubic  terms. $c_k\equiv 0$ for N=2. For $N\ge 3$ it has yet to be computed. 
\item $V_U/g_3^2N$ generates only terms in $b_k$. 
\item $V_Z/g^2_3N$ generates terms both in $b_k$ and for $N\ge 4$ in $z_k$.
\end{itemize}

\section{Conclusions}
 Using the full gauge independent effective action for {\it any} number of colours N Casimir scaling and universality leave two quantities to be matched: b and ${v\over {NT}}$. 
  So for {\it all} N $b=1.60379.., \mbox{~and} {v\over{NT}}=1.000811..$ from the N=2 calculation in ref.~\cite{deforkurk}.
 The coefficients  
$b_k$, and $z_k$  have been computed to one loop order for N=2,3,4~\cite{korthals}. 
The $c_k$ are yet to be computed for $N\ge 3$.  The SU(2) (SU(3)) action has  5 (9) independent couplings and 4 (5) constraints. Higher loop matching may fix more coefficients~\cite{york}.  Lattice computations have for $N\ge 5$ less numerical control (non-superrenormalizability). The number of terms in $V_Z$ grows like the number of partitions of N.



\end{document}